\documentstyle [12pt] {article}
\title{A SIMPLE DETERMINATION OF THE THERMODYNAMICAL CHARACTERISTICS OF THE WEAKLY CHARGED,  VERY THIN BLACK RING}

\author{Vladan Pankovi\'c$^{\ast,\sharp}$,
Simo Ciganovi\'c$^\sharp$\\
$^\ast$Department of Physics, Faculty of Sciences, 21000 Novi
Sad,\\ Trg Dositeja Obradovi\'ca 4. , Serbia, vdpan@neobee.net \\
$^\sharp$Gimnazija, 22320 Indjija, Trg Slobode 2a, Serbia \\}

\date {}
\begin{document}
\maketitle


\begin {abstract}
In our previous work we suggested a very simple, approximate
formalism for description of some basic (especially
thermodynamical) characteristics of a non-charged, rotating, very
thin black ring. Here, in our new work, generalizing our previous
results, we suggest a very simple, approximate description of some
basic (especially thermodynamical) characteristics of a weakly
charged, rotating, very thin black ring. (Our formalism is not
theoretically dubious, since, at it is not hard to see, it can
represent an extreme simplification of a more accurate, e.g.
Copeland-Lahiri, string formalism for the black hole description.)
Even if suggested formalism is, generally speaking,
phenomenological and rough, obtained final results, unexpectedly,
are non-trivial. Concretely, given formalism reproduces exactly
Bekenstein-Hawking entropy, Bekenstein quantization of the entropy
or horizon area and Hawking temperature of a weakly charged,
rotating, very thin black ring originally obtained earlier using
more accurate analysis by Emparan, Aestefanesei, Radu etc.
(Conceptually it is similar to situation in Bohr's atomic model
where energy levels are determined practically exactly even if
electron motion is described roughly.) Our formalism is physically
based on the assumption that circumference of the horizon tube
holds the natural (integer) number of corresponding reduced
Compton's wave length. (It is conceptually similar to Bohr's
quantization postulate in Bohr's atomic model interpreted by de
Broglie relation.) Also, we use, mathematically, practically only
simple algebraic equations.
\end {abstract}
\vspace {1.5cm}

In our previous work [1] we suggested a very simple, approximate
formalism for description of some basic (especially
thermodynamical) characteristics of a non-charged, rotating, very
thin black ring. Here, in our new work, generalizing our previous
results, we suggest a very simple, approximate description of some
basic (especially thermodynamical) characteristics of a weakly
charged, rotating, very thin black ring. (Our formalism is not
theoretically dubious, since, at it is not hard to see, it can
represent an extreme simplification of a more accurate, e.g.
Copeland-Lahiri [2], string formalism for the black hole
description.) Even if suggested formalism is, generally speaking,
phenomenological and rough, obtained final results, unexpectedly,
are non-trivial. Concretely, given formalism reproduces exactly
Bekenstein-Hawking entropy, Bekenstein quantization of the entropy
or horizon area and Hawking temperature of a weakly charged,
rotating, very thin black ring originally obtained earlier using
more accurate analysis by Emparan [3], Aestefanesei, Radu [4] etc.
(Conceptually it is similar to situation in Bohr's atomic model
where energy levels are determined practically exactly even if
electron motion is described roughly.) Our formalism is physically
based on the assumption that circumference of the horizon tube
holds the natural (integer) number of corresponding reduced
Compton's wave length. (It is conceptually similar to Bohr's
quantization postulate in Bohr's atomic model interpreted by de
Broglie relation.) Also, we use, mathematically, practically only
simple algebraic equations (by determination of Hawking
temperature we use additionally only simple differentiation of
Smarr relation).

As it is well-known [3], [5], [6], horizon of a rotating, very
thin black ring can be approximated by a torus with great radius
(i.e. distance from the center of the tube to the center of the
torus)
\begin {equation}
    R_{1} = R(\frac {\lambda}{\nu})^{\frac {1}{2}}
\end {equation}
and small radius (i.e. radius of the torus tube)
\begin {equation}
   R_{2} = R\nu
\end {equation}
, where $R$, $\lambda$ and $\nu$ represent corresponding variables
so that, for rotating, very thin black ring, it is satisfied
\begin {equation}
    0 < \nu < \lambda \ll 1     .
\end {equation}

Further, it is well-known [3], [4], that a weakly charged,
rotating very thin black ring holds additional two variables, $N$
(corresponding to dilaton coupling), and, $\mu$ (corresponding to
local charge $Q$) that satisfy conditions
\begin {equation}
    0 < N \leq 3
\end {equation}
\begin {equation}
    0 < µ \ll   1  .
\end {equation}

Finally, it is well-known too [3], [4] that weakly charged,
rotating, very thin black ring holds (in the natural units system
$\hbar = G = c = k = 1$) the following mass - $M$, angular
momentum - $J$, angular velocity - $\Omega$, horizon area - $A$,
entropy - $S$, temperature - $T$, local (electric) charge - Q, and
(electric) potential - F
\begin {equation}
    M = \frac {3\pi}{4}R^{2}(\lambda + \frac {N \mu}{3}) =
    \frac {3\pi}{4}  R_{1}R_{2} (\frac {\lambda}{\nu})^{\frac {1}{2}} (1+ \frac {N \mu}{3\lambda })
\end {equation}
\begin {equation}
    J = \frac {\pi R^{3}}{2} \lambda (1 - \frac {\nu}{\lambda}) ^{\frac {1}{2}}=
    \frac {\pi}{2} R^{2}_{1}R_{2} ( 1 -  \frac {\nu}{\lambda}) ^{\frac {1}{2}}
\end {equation}

\begin {equation}
   \Omega = \frac {1}{R}(1 - \frac {\nu}{\lambda}) ^{\frac {1}{2}}  =
   \frac {1}{R_{1}}( \frac {\lambda}{\nu}-1) ^{\frac {1}{2}}
\end {equation}
\begin {equation}
   A = 8 \pi^{2}R^{3}\nu^{\frac {3-N}{2}}\lambda^{\frac {1}{2}}(\mu + \nu)^{\frac {N}{2}} =
   8 \pi^{2} R^{2}_{2}R_{1} \nu^{\frac {-N}{2}}(\mu + \nu)^{\frac {N}{2}}
\end {equation}
\begin {equation}
    S = \frac {A}{4}= 2\pi^{2} R^{2}_{2}R_{1} \nu^{\frac {-N}{2}}(\mu + \nu)^{\frac {N}{2}}
\end {equation}
\begin {equation}
   T = \frac {1}{4\pi R}\nu^{\frac {N-1}{2}}\lambda^{-\frac {1}{2}}(\mu + \nu)^{\frac {-N}{2}} =
   \frac {1}{4\pi R_{2}} (\frac {\nu}{\lambda})^{\frac {1}{2}}\nu^{\frac {N}{2}} (\mu + \nu)^{\frac {-N}{2}}
\end {equation}
\begin {equation}
     Q = R N^{\frac {1}{2}} \mu^{\frac {1}{2}} =
     R_{1} (\frac {\nu}{\lambda}) ^{\frac {1}{2} }N^{\frac {1}{2}}\mu^{\frac
     {1}{2}}
\end {equation}
\begin {equation}
     \Phi = \frac {\pi}{2}R N^{\frac {1}{2}}\mu^{\frac {1}{2}}(\mu + \nu)^{\frac {-1}{2}} =
     \frac {\pi}{2}R_{2} N^{\frac {1}{2}}\mu^{\frac {1}{2}}(\mu + \nu)^{\frac {-1}{2}}\nu^{-1}  .
\end {equation}

Now, as well as it has been done in our previous work [1], {\it
suppose (postulate) } the following expression
\begin {equation}
      m_{+n}R_{2} = \frac {n}{2\pi}, \hspace{1cm}   {\rm for}  \hspace{1cm}  n=1,
      2,...
\end {equation}
where $ m_{+n}$ for $n = 1, 2, …$ represent masses of some small
quantum system captured at given weakly charged, rotating, very
thin black ring horizon. It implies
\begin {equation}
      2\pi R_{2} = \frac {n}{m_{+n}} = n \lambda_{r+n}   , \hspace{1cm}   {\rm for}  \hspace{1cm}  n=1,
      2,...         .
\end {equation}
Here, obviously, $2\pi R_{2}$ represents the circumference of
small circle at given weakly charged,  rotating, very thin black
ring horizon, while
\begin {equation}
      \lambda_{r+n}=\frac {1}{m_{+n}}
\end {equation}
represents the $n$-th reduced Compton wavelength of a quantum
system with mass $ m_{+n}$ captured at weakly charged, very thin
black ring horizon for $n = 1, 2, …$ .

Expression (15) simply means that {\it circumference of given
small circle (tube) at given weakly charged, rotating, very thin
black ring horizon holds exactly n corresponding} $n$-{\it th
reduced Compton wave lengths of mentioned small quantum system
for} $n = 1, 2, …$ . (Obviously, it is conceptually similar to
well-known Bohr's quantization postulate interpreted by de Broglie
relation (according to which circumference of $n$-th electron
circular orbit contains exactly $n$ corresponding $n$-th de
Broglie wave lengths, for $n = 1, 2, …$).

According to (14) it follows
\begin {equation}
       m_{+n}= \frac {n}{2\pi R_{2}} = n m_{+1} , \hspace{1cm}   {\rm for}  \hspace{1cm}  n=1,
      2,...
\end {equation}
where
\begin {equation}
       m_{+1} = \frac {1}{2\pi R_{2}}          .
\end {equation}

Now, analogously to procedure in [1], {\it suppose (postulate)
that given weakly charged,  rotating, very thin black ring entropy
is proportional to quotient of given weakly charged,  rotating,
very thin black ring mass} $M$ {\it and minimal mass of mentioned
small quantum system} $ m_{+1}$. More precisely, suppose
(postulate)
\begin {equation}
      S = \gamma \frac{M}{ m_{+1}} =
      \gamma (\frac {3 \pi^{2}}{2}) R_{1}R^{2}_{2}(\frac {\lambda}{\nu})^{\frac {1}{2}} (1 + \frac {N\mu}{3\lambda})
\end {equation}
where $\gamma$ represents some correction factor, i.e. parameter
that can be determined by comparison of (19) and (10). It yields
\begin {equation}
      \gamma = \frac {4}{3}(\frac {\nu}{\lambda})^{\frac {1}{2}}(\mu + \nu)^{\frac {N}{2}}\nu^{\frac {-N}{2}}(1 + \frac {N\mu}{3\lambda})^{-1}                         .
\end {equation}

Also, we shall define normalized mass of given weakly charged,
rotating, very thin black ring
\begin {equation}
      \tilde {M} = \gamma M = \pi R_{1}R_{2}(\mu + \nu)^{\frac {N}{2}}\nu^{\frac {-N}{2}}     .
\end {equation}

Further, we shall differentiate (9), (10) supposing approximately
that $R_{2}$  represents unique variable while $ R_{1}$, $\nu$,
$\lambda$, $\mu$ and $N$ can be considered as constant parameters.
It yields
\begin {equation}
      dA = 16 \pi^{2}R_{1} R_{2}(\mu + \nu)^{\frac {N}{2}}\nu^{\frac {-N}{2}} d R_{2}
\end {equation}
\begin {equation}
      dS = 4\pi^{2}R_{1} R_{2}(\mu + \nu)^{\frac {N}{2}}\nu^{\frac {-N}{2}} d R_{2}               .
\end {equation}

Since, according to (21),
\begin {equation}
      d\tilde {M} = \pi R_{1}(\mu + \nu)^{\frac {N}{2}}\nu^{\frac {-N}{2}}dR_{2}
\end {equation}
that yields
\begin {equation}
      dR_{2} = \frac { d\tilde {M}}{\pi R_{1}(\mu + \nu)^{\frac {N}{2}}\nu^{\frac {-N}{2}}}
\end {equation}
, (22), (23) turn out in
\begin {equation}
      dA = 16\pi R_{2} d\tilde {M}
\end {equation}
\begin {equation}
      dS = 4 p \pi R_{2} d\tilde {M}               .
\end {equation}

Given expressions can be approximated by the following finite
difference forms
\begin {equation}
      \Delta A = 16\pi R_{2} \Delta \tilde {M}
      \hspace{1cm} {\rm for}  \hspace{0.5cm} \Delta \tilde {M}\ll  \tilde {M}
\end {equation}
\begin {equation}
      \Delta S = 4 \pi R_{2} \Delta \tilde {M}
      \hspace{1cm} {\rm for}  \hspace{0.5cm} \Delta \tilde {M}\ll  \tilde {M}
\end {equation}
Further, we shall assume
\begin {equation}
   \Delta \tilde {M}= n m_{+1} \hspace{1cm}   {\rm for}  \hspace{1cm}  n=1,
      2,...  .
\end {equation}
Introduction of (30) in (28), (29), according to (18), yields
\begin {equation}
    \Delta A = 8n = 2n (2)^{2} \hspace{1cm}   {\rm for}  \hspace{1cm}  n=1,
      2,...
\end {equation}
\begin {equation}
    \Delta S = 2n  \hspace{1cm}   {\rm for}  \hspace{1cm}  n=1,
      2,...
\end {equation}
that exactly represent Bekenstein quantization of the black hole
horizon surface (where $(2)^{2}$ represents the surface of the
quadrate whose side length represents twice Planck length, i.e. 1)
and entropy.

Now, we shall determine Hawking temperature of given weakly
charged, rotating, very thin black ring in our approximation in
the following way. We shall start from the first thermodynamical
law for charged black rings
\begin {equation}
   dM = TdS +  \Omega dJ + Qd\Phi                .
\end {equation}

Further, we shall introduce an approximation, opposite to
approximation introduced previously by deduction of Bekenstein
entropy or horizon area quantization. Namely, we shall again
suppose approximately that in (33) known functions $M$ (6), $J$
(7), $\Omega$ (8), $S$ (10), $Q$ (12) and $\Phi$ (13) as well as
unknown function $T$ that will be determined by (33) represent
functions of only one variable $R_{2}$ while $ R_{1}$, $\nu$,
$\lambda$, $\mu$ and $N$ can be considered as constant parameters.
Secondly, it can be supposed that under first supposition
expression (33) cannot be exactly conserved so that given
expression within given approximation must be corrected by two
additional correction factors $\alpha$ and $\beta$ in the
following way
\begin {equation}
   \alpha dM = TdS +  \Omega dJ + \beta Qd \Phi                .
\end {equation}
Given correction factors $\alpha$ and $\beta$ will be determined
by comparison of the obtained value of $T$ with exact expression
(11).

Application of the introduced approximation rule and (6)-(10),
(12), (13) on (34) yields
\begin {equation}
   \alpha \frac {3\pi}{4}R_{1} (\frac {\lambda}{\nu})^{\frac {1}{2}}(1 + \frac {N \mu}{3\lambda})dR_{2} =
\end {equation}
\[T4 \pi^{2}R_{1}R_{2}\nu^{\frac {-N}{2}}(\mu + \nu)^{\frac
{N}{2}}dR_{1}
 + \frac {\nu}{R_{2}}( 1 - \frac {\nu}{\lambda})^{\frac {1}{2}}\pi  R_{1} R_{2}\nu^{-1}
 (\frac {\lambda}{\nu}  -1)^{\frac {1}{2}}dR_{2} \]
 \[  + \beta R_{2}\frac {1}{\nu} N^{\frac {1}{2}}\mu^{\frac {1}{2}}(\mu + \nu) ^{\frac {1}{2}}
\frac {\pi}{2} (\frac {\nu}{\lambda})^{\frac {1}{2}}N^{\frac
{1}{2}}\mu^{\frac {1}{2}}(\mu + \nu) ^{\frac {-1}{2}}dR_{1} \]

which, after simple transformations, yields
\begin {equation}
  T = \frac {1}{4\pi R_{2}} (\frac {\nu}{\lambda})^{\frac {1}{2}} \nu^{\frac {N}{2}}
  (\mu + \nu)^{\frac {-N}{2}} (\frac {3\alpha}{4} \frac {\lambda}{\nu}   + \frac {\alpha}{4}\frac {N \mu}{\nu} - \frac {1}{2}\frac {\lambda}{\nu}+ \frac {1}{2} - \frac {\beta}{2} \frac {N \mu}{\nu}).
\end {equation}

Now, comparison of (36) and (11) implies
\begin {equation}
  \alpha \frac {3}{4}\frac {\lambda}{\nu}   - \frac {1}{2}  \frac {\lambda}{\nu}= 0
\end {equation}
\begin {equation}
 (\frac {\alpha}{4}) \frac {N\mu }{ \nu}  - (\frac {\beta}{2}) \frac {N\mu }{ \nu}= 0
\end {equation}
that implies
\begin {equation}
   \alpha  =  \frac {2}{3}
\end {equation}
\begin {equation}
   \beta  =  \frac {1}{3}                                          .
\end {equation}
It, on the one hand, means that (36) turns out in
\begin {equation}
  T = \frac {1}{2}\frac {1}{4\pi R_{2}} (\frac {\nu}{\lambda})^{\frac {1}{2}} \nu^{\frac {N}{2}} (\mu + \nu)^{\frac {-N}{2}}
\end {equation}
that represents one half of the exact Hawking temperature (11). It
can be considered as a satisfactory result, even if there is no
complete equivalence between exact, (11), and approximate, (41),
term.
    On the other hand, introduction of the correction factors (39), (40), in the corrected first thermodynamical law (34) yields
\begin {equation}
   \frac {2}{3}dM = TdS +  \Omega dJ + \frac {1}{3}Qd\Phi
\end {equation}
that, obviously, corresponds to Smarr relation  [2], [3] exactly
satisfied for black rings (including weakly charged, rotating,
very thin black ring too)
\begin {equation}
   \frac {2}{3}M = TS +  \Omega J + \frac {1}{3}Q\Phi                       .
\end {equation}
It, practically, implies that correction coefficients $\alpha$ and
$\beta$  can be determined by Smarr relation. In other word,
within our approximative method, Hawking temperature can be
determined exactly, starting from Smarr relation instead of first
thermodynamical law.

In this way we have reproduced, i.e. determined simply,
approximately three most important thermodynamical characteristics
of a weakly charged, rotating, very thin black ring:
Bekenstein-Hawking entropy, Bekenstein quantization of the horizon
area or entropy, and, Hawking temperature . It can be observed,
roughly speaking, that all these characteristics are caused by
standing (reduced Compton) waves at small circles (torus tube
circles) at horizon area only. In other words, within given
approximation as well as within a more accurate analysis,
thermodynamical characteristics of given rotating, very thin black
ring are practically independent of the great circle of the torus.

In conclusion it can be shortly repeated and pointed out the
following. In this work we suggested a simple, approximate
formalism for description of some basic (especially
thermodynamical) characteristics of a weakly charged, rotating,
very thin black ring. (In fact, our formalism is not theoretically
dubious, since, at it is not hard to see, it can correspond to an
extreme simplification of a more accurate, Copeland-Lahiri string
formalism for the black hole description.) Even if suggested
formalism is, generally speaking, phenomenological and rough,
obtained final results, unexpectedly, are non-trivial. Concretely,
given formalism reproduces exactly Bekenstein-Hawking entropy,
Bekenstein quantization of the entropy or horizon area and Hawking
temperature of a rotating, very thin black ring obtained earlier
using more accurate analysis (Emparan, Aestefanesei, Radu etc.).
Our formalism is physically based on the assumption that
circumference of the horizon tube holds the natural (integer)
number of corresponding reduced Compton's wave length. (It is
conceptually similar to Bohr's quantization postulate in Bohr's
atomic model interpreted by de Broglie relation.) Also, we use,
mathematically, practically only simple algebraic equations.

\section {References}

\begin {itemize}
\item [[1]] V. Pankovi\'c, S Ciganovi\'c, {\it A Simple Determination of the Thermodynamical Characteristics of a Very Thin Black Ring}, gr-qc/0808.0618
\item [[2]] E. J. Copeland, A.Lahiri, Class. Quant. Grav. , {\bf 12} (1995) L113 ; gr-qc/9508031
\item [[3]] R. Emparan, {\it Rotating Circular Strings and Infinite Non-Uniqueness of Black Rings}, hep-th/0402149
\item [[4]] D. Aestefanesei, E. Radu, {\it Quasilocal Formalism and Black ring Thermodynamics}, hep-th/0509144
\item [[5]] R. Emparan, H. S. Reall, {\it A Rotating Black Ring in Five Dimensions}, hep-th/0110260
\item [[6]] H. Elvang, R. Emparan, A. Virmani, {\it Dynamics and Stability of Black Rings}, hep-th/0608076

\end {itemize}

\end {document}